\documentclass[a4paper,10pt]{article}

\usepackage{graphicx}             
\DeclareGraphicsExtensions{.ps,.eps,.pstex}

\usepackage[latin1]{inputenc}     %

\usepackage{epsfig}               
\psfull                           
\usepackage{color}                
\usepackage{colortbl}

\usepackage{amsmath}              
\usepackage{bbm}                  

\usepackage[sf,bf]{caption}       

\usepackage{sectsty}              
\sectionfont{\sffamily}
\subsectionfont{\sffamily}

\newcommand{\op}[1]{\mathcal{#1}}

\newcommand{\ket}[1]{\left| #1 \right>}
\newcommand{\bra}[1]{\left< #1 \right|}

\newcommand{\bret}[1]{\left< #1 \right>}

\newcommand{\e}{\mathrm{e}}

\newcommand{\di}{\mathrm{d}}

\newcommand{\I}{\mathrm{i}}
\newcommand{\Tr}{\mathrm{Tr}}

\newcommand{\slim}{\mskip 1.5mu}

\newcommand{\eVdist}{\kern-0.06667em}

\newcommand{\Mev}{{\text{Me}\eVdist\text{V\/}}}
\newcommand{\Gev}{{\text{Ge}\eVdist\text{V\/}}}

\newcommand{\mev}{{\,\text{Me}\eVdist\text{V\/}}}
\newcommand{\gev}{{\,\text{Ge}\eVdist\text{V\/}}}

\newcommand{\tsi}{t_{\text{sink}}}
\newcommand{\tsc}{t_{\text{source}}}
\newcommand{\tsh}{t_{\text{shift}}}

\title{\sffamily\bfseries{The pion form factor from lattice QCD\\ with two
    dynamical flavours}\\[-1ex]%
  \begin{picture}(0,0)(0,0)%
    \put(0,102.5){\makebox(0,0)[l]{\textnormal{\normalsize DESY 06-135}}}%
    \put(0,90){\makebox(0,0)[l]{\textnormal{\normalsize Edinburgh 2006/17}}}%
    \put(0,79){\makebox(0,0)[l]{\textnormal{\normalsize LTH 710}}}%
    \put(0,66){\makebox(0,0)[l]{\textnormal{\normalsize TUM/T39-06-06}}}
  \end{picture}%
  \hrulefill}

\author{%
  D.~Br\"ommel$^{\,a,b}$, %
  M.~Diehl$^{a}$, %
  M.~G\"ockeler$^b$, %
  Ph.~H\"agler$^c$, \\ %
  R.~Horsley$^d$, %
  Y.~Nakamura$^e$, %
  D.~Pleiter$^e$, %
  P.E.L.~Rakow$^f$,\\ %
  A.~Sch\"afer$^b$, 
  G.~Schierholz$^{a,e}$, %
  H.~St\"uben$^g$ %
  and J.M.~Zanotti$^d$\\[2ex]%
  {\begin{minipage}{12cm}
    \small\textit{\flushleft{
    $^a$ Theory Group, Deutsches Elektronen-Synchrotron~DESY,
    22603~Hamburg, \phantom{$^a$ }Germany\\[.3ex]
    $^b$ Institut f\"ur Theoretische Physik, Universit\"at Regensburg,
    93040~Regensburg, \phantom{$^b$ }Germany\\[.3ex]
    $^c$ Institut f\"ur Theoretische Physik T39, Physik-Department der TU
    M\"unchen, \phantom{$^c$ }85747~Garching, Germany\\[.3ex]
    $^d$ School of Physics, University of Edinburgh, Edinburgh EH9 3JH,
    UK\\[.3ex]
    $^e$ John von Neumann-Institut f\"ur Computing NIC~/~DESY,
    15738~Zeuthen, \phantom{$^e$ }Germany\\[.3ex]
    $^f$ Theoretical Physics Division, Dep.\ of Math.\ Sciences,
    University of Liverpool, \phantom{$^f$ }Liverpool L69~3BX, UK\\[.3ex]
    $^g$ Konrad-Zuse-Institut f\"ur Informationstechnik ZIB, 14195~Berlin,
    Germany\\}}
  \end{minipage}}%
  \vspace*{2.5ex}\\%
  QCDSF/UKQCD Collaboration
}

\date{June 2007}

\begin{document}

\maketitle

\renewcommand\abstractname{\sffamily{Abstract}}
\begin{abstract}
  We compute the electromagnetic form factor of the pion using
  non-perturbatively $O(a)$ improved Wilson fermions. The calculations are done
  for a wide range of pion masses and lattice spacings. We check for finite size
  effects by repeating some of the measurements on smaller lattices. The large
  number of lattice parameters we use allows us to extrapolate to the physical
  point. For the square of the charge radius we find $\bret{r^2} =
  0.441(19)~\text{fm}^2$, in good agreement with experiment.
\end{abstract}

\section{Introduction\label{sec:intro}}

For some time now it has been possible to explore the structure of hadrons from
first principles using lattice QCD.  Since the pion is the lightest QCD bound
state and plays a central role in chiral symmetry breaking and in low-energy
dynamics, a thorough investigation of its internal structure in terms of quark
and gluon degrees of freedom should be particularly interesting.  We have
started to explore the structure of the pion in a framework using generalised
parton distributions, or more precisely their moments \cite{Brommel:2005ee}. As
a generalisation of parton distributions and form factors they contain both as
limiting cases. In this work we restrict ourselves to results for the pion
electromagnetic form factor $F_{\pi}$ from $N_f=2$ lattice QCD simulations,
based on $O(a)$ improved Wilson fermions and Wilson glue. Initial studies on the
pion form factor by Martinelli {\it{et al.\/}} and Draper {\it{et al.\/}}
\cite{Martinelli:1987bh,Draper:1988bp} were followed by recent
simulations in quenched \cite{Capitani:2005ce,vanderHeide:2003kh,%
  Abdel-Rehim:2004gx,Nemoto:2003ng} and unquenched QCD
\cite{Bonnet:2004fr,Hashimoto:2005am}. In this work, we improve upon previous
calculations by extracting the pion form factor for a much larger number of
$\beta$, $\kappa$ combinations, which allows us to study both the chiral and the
continuum limit. Furthermore, two finite size runs make estimates of the volume
effect possible.

\section{The pion form factor in lattice QCD\label{sec:lat}}

The pion electromagnetic form factor $F_{\pi}$ describes how the vector current
\begin{equation}
  \label{eq:def:current}
  V_{\mu}(x)=\textstyle{\frac{2}{3}}\, \overline{u}(x) \gamma_{\mu} u(x) - 
  \textstyle{\frac{1}{3}}\, \overline{d}(x) \gamma_{\mu} d(x)
\end{equation}
couples to the pion. Writing $p$ and $p'$ for the incoming and outgoing momenta
of the pion, it is defined by
\begin{equation}
  \label{eq:def:fpi}
  \bra{\pi^{+}(p')} V_{\mu}(0) \ket{\pi^{+}(p)} =
  (p'_{\mu}+p^{\phantom{'}}_{\mu})\, F_{\pi}(Q^2)\,,
\end{equation}
where the momentum transfer is $q_{\mu} = (p'_{\mu} - p^{\phantom{'}}_{\mu})$
and its invariant square is $q^2=-Q^2$.

For our lattice calculation we want to simplify the flavour structure of
Eq.~\eqref{eq:def:current}. Invoking isospin symmetry one finds
\begin{equation}
  \label{eq:isospin}
  \bra{\pi^+} \textstyle{\frac{2}{3}}\overline{u} \gamma_{\mu} u - 
  \textstyle{\frac{1}{3}} \overline{d} \gamma_{\mu} d \ket{\pi^+} =\\
  \bra{\pi^+} \overline{u} \gamma_{\mu} u \ket{\pi^+} = - \bra{\pi^+}
  \overline{d} \gamma_{\mu} d \ket{\pi^+} \,.
\end{equation}
It is hence sufficient to limit the calculation to a single quark flavour in the
vector operator. We use the unimproved local vector current on the lattice; the
corrections due to the improvement term \cite{Sint:1997jx} are quite small and
will be discussed later. Since this current is not conserved, renormalisation
has to be taken into account. Because in the forward limit ($Q^2=0$) the form
factor is simply the electric charge of the pion, we can normalise our data
appropriately. We can also use the known renormalisation constant $Z_V$ (taken
for example from \cite{Bakeyev:2003ff}) as a cross-check for our simulation.

\begin{figure}
  \begin{center}
    \begin{picture}(0,0)
      \includegraphics[width=4cm]{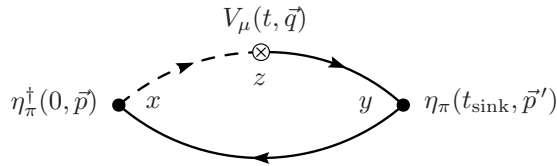}
    \end{picture}
    \setlength{\unitlength}{0.01cm}
    \begin{picture}(400,220)(0,0)
      \put(-95,100){\makebox(0,0)[c]{$\eta_{\pi}^{\dagger}(0,\vec{p})$}}%
      \put(182,208){\makebox(0,0)[c]{$V_{\mu}(t,\vec{q})$}}%
      \put(480,100){\makebox(0,0)[c]{$\eta_{\pi}(\tsi,\vec{p}\,')$}}%
      \put(35,98.5){\makebox(0,0)[c]{$x$}}%
      \put(175,130){\makebox(0,0)[c]{$z$}}%
      \put(315, 95){\makebox(0,0)[c]{$y$}}%
    \end{picture}
  \end{center}\vspace*{-1em}
  \caption{\label{fig:3pt}A sketch of the three-point function with
    the pion source at time 0, pion sink at $\tsi$, and the operator
    acting at time $t$.}
\end{figure}

To compute the matrix elements in Eq.~\eqref{eq:def:fpi} on the lattice, one has
to evaluate pion three-point and two-point functions.  We then apply a standard
procedure to extract the pion form factor $F_{\pi}$, where one constructs an
appropriate ratio for the observable \cite{Capitani:1998ff,Gockeler:2003ay}. Let
us start by looking at the three-point function. The general form is given by
the correlation function
\begin{equation}
  \label{eq:3pt:gen}
  C_{\text{3pt}}(t,\vec{p}\,',\vec{p}\,) =
  \bret{ {\eta_{\pi}^{\phantom{\dagger}}(\tsi,\vec{p}\,') \,\;\overline{u}(t)
  \gamma_{\mu} u(t) \,\; 
    \eta_{\pi}^{\dagger}(\tsc,\vec{p}\,)}}
\end{equation}
and depicted in Fig.~\ref{fig:3pt}. Here we denote the sink and source operators
for a pion with given momentum and at given time-slice by
$\eta_{\pi}(\tsi,\vec{p}\,')$ and $\eta_{\pi}^{\dagger}(\tsc,\vec{p}\,)$,
respectively. Using the transfer matrix formalism and inserting complete sets of
energy eigenstates, the three-point function is then of the form
\begin{multline}
  \label{eq:3pt:hil}
  C_{\text{3pt}}(t,\vec{p}\,',\vec{p}\,) = 
     \bra{\pi(\vec{p}\,')} \overline{u}(0)
  \gamma_{\mu} u(0) \ket{\pi(\vec{p}\,)}
  \frac{\bra{0} \eta_{\pi}(\vec{p}\,') \ket{\pi(\vec{p}\,')}
   \bra{\pi(\vec{p}\,)}
   \eta_{\pi}^{\dagger}(\vec{p}\,) \ket{0}}{2E_{p'}\,2E_{p}} \\
   \times \left( \text{e}^{-E_{p'}(\tsi-t)-E_{p}\,t} + (-1)^{n_4}\,
    \text{e}^{-E_{p'}(t-\tsi)-E_{p}\,(T-t)}
     \right) 
+\cdots\,,
\end{multline}
where $T$ is the time extent of our lattice and
\begin{equation}
  \label{eq:n4:def}
n_4=\left\{
  \begin{array}{l}
    1\text{~~for~}\mu = 4\,,\\
    0\text{~~otherwise.}
  \end{array}\right.
\end{equation}
Note that we have omitted excited states in Eq.~\eqref{eq:3pt:hil} and already
inserted our choice for the time-slice of the pion source, $\tsc=0$. We choose
the sink of the three-point function as $\tsi=T/2$, so that the correlation
function is symmetric or antisymmetric with respect to this time,
\begin{equation}
  \label{eq:sign}
  C_{\text{3pt}}(t,\vec{p}\,',\vec{p}\,) = (-1)^{n_4}\,
  C_{\text{3pt}}(T-t,\vec{p}\,',\vec{p}\,) .
\end{equation}
We can then separate the correlation function into contributions from $t$ to the
left and to the right of $\tsi$ (referred to as l.h.s.\ and r.h.s.\ in the
following) and neglect either the second or first term in Eq.~\eqref{eq:3pt:hil}
since it is exponentially suppressed in the regions of $t$ from which we will
extract the form factor.

The two-point function has the form
\begin{equation}
  \label{eq:2pt}
  C_{\text{2pt}}(t,\vec{p}\,) = \frac{\bra{0} \eta_{\pi}(\vec{p}\,)
    \ket{\pi(\vec{p}\,)} \bra{\pi(\vec{p}\,)} \eta_{\pi}^{\dagger}(\vec{p}\,)
    \ket{0}} {2E_{p}} \text{e}^{-E_{p}T/2} \, 2\cosh [ E_{p}(T/2-t) ]
    +\cdots \,,
\end{equation}
where again we omitted higher energy states. Comparing the two- and three-point
functions \eqref{eq:2pt} and \eqref{eq:3pt:hil}, a ratio can be constructed that
eliminates the overlap factors such as $\bra{0} \eta_{\pi}(\vec{p}\,')
\ket{\pi(\vec{p}\,')}$ and partially cancels the exponential time behaviour
appearing in Eq.~\eqref{eq:3pt:hil}. This technique also has the advantage that
fluctuations of the correlation functions tend to cancel in the ratio and we
thus obtain a better signal. With our choice $\tsi = T/2$, such a ratio is
\begin{equation}
  \label{eq:ratio:lat}
  R(t) = %
  \frac{C_{\text{3pt}}(t,\vec{p}\,',\vec{p}\,)}
  {C_{\text{2pt}}(\tsi,\vec{p}\,')} \; \left[ %
    \frac{C_{\text{2pt}}(\tsi-t,\vec{p}\,)\, C_{\text{2pt}}(t,\vec{p}\,')\,
      C_{\text{2pt}}(\tsi,\vec{p}\,')}%
    {C_{\text{2pt}}(\tsi-t,\vec{p}\,')\, C_{\text{2pt}}(t,\vec{p}\,)\,
      C_{\text{2pt}}(\tsi,\vec{p}\,)}%
  \right]^{\!\frac{1}{2}}\,.
\end{equation}
Similar ratios have already been used in earlier works on pion and nucleon
structure. Here we take the somewhat more complicated ratio
\eqref{eq:ratio:lat}, which was used for the nucleon in \cite{Capitani:1998ff},
because we use momentum combinations with $|\vec{p}\,| \neq |\vec{p}\,'|$.
Contributions to this ratio from excited states with energy $E'$ are suppressed
as long as $\tsi-t \gg 1/(E'-E)$ and $t \gg 1/(E'-E)$ where $E$ is the pion
energy. A potential problem is that, due to the exponential decay of the pion
two-point function, the signal at $t=\tsi$ for non-vanishing momenta is poor.
For finite statistics the two-point function can then take negative values,
which prevents one from evaluating the square root.  We try to overcome this
difficulty by shifting the two-point functions $C_{\text{2pt}}(t,\vec{p}\,)$
that enter with $t=\tsi$.  Using the identity
\begin{equation}
  \label{eq:ratio:id}
  C_{\text{2pt}}(\tsi,\vec{p}) =
    \frac{C_{\text{2pt}}(\tsi-\tsh,\vec{p})}{\cosh(E_{p}\, 
    \tsh)}
\end{equation}
valid for $\tsi=T/2$, we shift by $\tsh=6$, which significantly reduces the
number of negative two-point functions. Nevertheless there are still momentum
transfers $Q^2$ for which the argument of the square root in the ratio
\eqref{eq:ratio:lat} is negative. Those values are discarded when we evaluate
the form factor.

\begin{figure}[!t]
  \centering
  \includegraphics[angle=270]{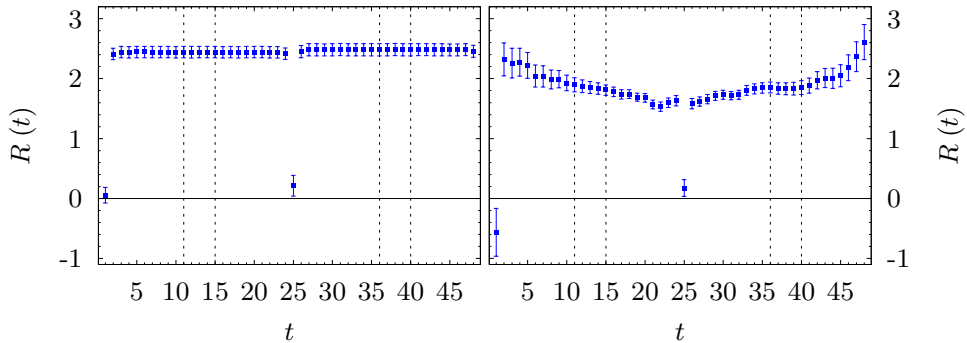}
  \caption{\label{fig:pp1} Examples of the ratio $R(t)$ in
    Eq.~\eqref{eq:ratio:lat} on the $24^3 \times 48$ lattice at $\beta=5.25$,
    $\kappa=0.13575$, multiplied with an appropriate sign factor $(-1)^{n_4}$
    for $t>\tsi$. The left plot shows a proper plateau in the forward case
    $Q^2=0$ and the right plot shows the ratio for $Q^2=0.31\gev^2$ where no
    plateau is expected (see text).
    The dashed lines indicate the regions we average over.}
\end{figure}

For $Q^2 \neq 0$ the ratio \eqref{eq:ratio:lat} does not exhibit a
proper plateau that could immediately be used for fitting. This is due to our
choice for $\tsi$, for which the time dependence of the pion two-point function
cannot be approximated by a single exponential in the $t$ regions we use to
extract the form factor, see Eq.~\eqref{eq:2pt}. In fact, we now show that the
ratio is approximately antisymmetric around the central point $t=\tsi/2=T/4$ of
the l.h.s.\ (as well as around $t=3T/4$ on the r.h.s.). Defining $\delta\equiv
t-\tsi/2$ and expanding the ratio and its exponentials in
Eq.~\eqref{eq:ratio:lat} around $\delta = 0$ we find
\begin{equation}
  \label{eq:expansion1}
  R(t)=C(E_{p},E_{p'},Q^2)\left[1+2 \delta\, c_\delta(E_{p},E_{p'})
    +2 \delta^2\, c_\delta^2(E_{p},E_{p'}) + \op{O}(\delta^3)\right]\,,
\end{equation}
where
\begin{eqnarray}
  \label{eq:expansion2}
  c_\delta(E_{p},E_{p'}) &=& \frac{E_{p'}}{1+\e^{E_{p}\,\tsi}} -
  \frac{E_{p}}{1+\e^{E_{p'}\,\tsi}}\,,\nonumber\\[1ex]
  C(E_{p},E_{p'},Q^2) &=& \frac{(p'_{\mu}+p^{\phantom{'}}_{\mu})} {4
  \sqrt{E_{p'}E_{p}}} \, F_{\pi}(Q^2)\,.
\end{eqnarray}
When averaging $R(t)$ in a symmetric interval around $t=T/4$, the antisymmetric
piece proportional to $c_\delta$ in \eqref{eq:expansion1} drops out. However,
such an averaged signal also includes unwanted symmetric contributions.
Fortunately, for our pion masses and lattice momenta already the leading
symmetric term is negligible, because with the lattice spacing $a$ we have
$c_\delta^2\sim 10^{-4}a^{-2}$ and $\delta^2 \le 4a^2$ in our fits. We hence
obtain a good signal for the averaged ratio. The same is true for the r.h.s.\ 
ratio and its central point $t=3T/4$. A typical ratio at non-zero momentum
transfer is shown in Fig.~\ref{fig:pp1} for one of our data sets, along with the
familiar plateau for zero momentum transfer. Note that the ratio
\eqref{eq:ratio:lat} does not exhibit a plateau for arbitrary momenta. To
visualise the absence of possible contributions from excited states one has to
consider the ratio for $|\vec{p}\,|=|\vec{p}\,'|$. In this case the time
dependence of the three-point function \eqref{eq:3pt:hil} should vanish. We have
checked that this is indeed the case in the region we average over, within the
expected increase of noise for higher momenta or lower pion masses.

From Eqs.~\eqref{eq:expansion1} and \eqref{eq:expansion2} we see that the
lattice ratio \eqref{eq:ratio:lat} can be used to extract the form factor
$F_\pi(Q^2)$. Using then several combinations of momenta $p$ and $p'$ that all
give the same $Q^2$ provides an over-constrained set of equations, from which we
determine $F_{\pi}(Q^2)$ by $\chi^2$ minimisation. We increase the quality of
our signal by averaging the ratio over the contributions on the l.h.s.\ and
r.h.s. This requires the additional sign factor $(-1)^{n_4}$ between the two
sides, as can be seen in Eq.~\eqref{eq:sign}. The energies $E_{p}$ and $E_{p'}$
appearing in \eqref{eq:ratio:id} and \eqref{eq:expansion2} are calculated using
the lattice pion masses and the continuum dispersion relation. We also performed
a test of the dispersion relation for some of our lattices. It was increasingly
difficult to extract a signal for higher momenta, especially for the lowest pion
masses. However, we found that the continuum dispersion relation can be used to
describe the data and that a lattice dispersion relation is not favoured.

\section{Simulation details\label{sec:sim}}

We perform our simulations with two flavours of non-perturbatively
clover-im\-proved dynamical Wilson fermions and Wilson glue. Using these
actions, the QCDSF and UKQCD collaborations have generated gauge field
configurations with the parameters given in Table~\ref{tab:lats}, where we have
used the Sommer parameter with $r_0=0.467$~fm (see \cite{Khan:2006de} and
\cite{Aubin:2004wf}) to set the physical scale. This large set of lattices
enables us to extrapolate to the chiral and the continuum limit. For two sets of
parameters ($\beta=5.29$, $\kappa=0.1355, 0.1359$) we also have a choice of
lattice volumes ($12^3\times32$, $16^3\times32$ and $24^3\times48$) in order to
study finite volume effects.
\begin{table}[bt]
  \caption{\label{tab:lats} Overview of our lattice parameters. For physical
    units the Sommer parameter with $r_0=0.467$~fm has been used. The error on
    $m_{\pi}$ is purely statistical.}
  \begin{center}
  \begin{tabular}{cccccccc}
    \hline\hline%
    $\beta$ & \# & $\kappa$ & $N^3\times T$ & $m_{\pi}$~[\Gev] & $a$~[fm] & 
    $L$~[fm]& $N_{\text{traj}}$\rule{0pt}{2.3ex}\rule[-1.2ex]{0pt}{2.5ex}\\%
    \hline
   5.20 & 1& 0.13420 & $16^3\times 32$ &1.007(2) & 0.115 & 1.8 &
   $\mathcal{O}(5000)$\rule{0pt}{2.4ex}\\
        & 2& 0.13500 & $16^3\times 32$ &0.833(3) & 0.098 & 1.6 &
        $\mathcal{O}(8000)$\\%
        & 3& 0.13550 & $16^3\times 32$ &0.619(3) & 0.093 & 1.5 &
        $\mathcal{O}(8000)$\\%
   \hline            
   5.25 & 4& 0.13460 & $16^3\times 32$ &0.987(2) & 0.099 & 1.6 &
   $\mathcal{O}(5800)$\rule{0pt}{2.4ex}\\%
        & 5& 0.13520 & $16^3\times 32$ &0.829(3) & 0.091 & 1.5 &
        $\mathcal{O}(8000)$\\%
        & 6& 0.13575 & $24^3\times 48$ &0.597(1) & 0.084 & 2.0 &
        $\mathcal{O}(5900)$\\%
   \hline            
   5.26 & 7& 0.13450 & $16^3\times 32$ &1.011(3) & 0.099 & 1.6 &
   $\mathcal{O}(4000)$\rule{0pt}{2.4ex}\\%
   \hline            
   5.29 & 8& 0.13400 & $16^3\times 32$ &1.173(2) & 0.097 & 1.6 &
   $\mathcal{O}(4000)$\rule{0pt}{2.4ex}\\%
        & 9& 0.13500 & $16^3\times 32$ &0.929(2) & 0.089 & 1.4 &
        $\mathcal{O}(5600)$\\%
        &10& 0.13550 & $24^3\times 48$ &0.769(2) & 0.084 & 2.0 &
        $\mathcal{O}(2000)$\\%
        &11& 0.13590 & $24^3\times 48$ &0.591(2) & 0.080 & 1.9 &
        $\mathcal{O}(5900)$\\%
        &12& 0.13620 & $24^3\times 48$ &0.400(3) & 0.077 & 1.9 &
        $\mathcal{O}(5600)$\\%
   \hline            
   5.40 &13& 0.13500 & $24^3\times 48$ &1.037(1) & 0.077 & 1.8 &
   $\mathcal{O}(3700)$\rule{0pt}{2.4ex}\\%
        &14& 0.13560 & $24^3\times 48$ &0.842(2) & 0.073 & 1.8 &
        $\mathcal{O}(3500)$\\%
        &15& 0.13610 & $24^3\times 48$ &0.626(2) & 0.070 & 1.7 &
        $\mathcal{O}(3900)$\\%
    \hline\hline
  \end{tabular}
  \end{center}
\end{table}

Starting with the lattice version of the three-point function,
Eq.~\eqref{eq:3pt:gen}, we follow \cite{Best:1997qp} and find that it is
sufficient to calculate
\begin{equation}
  \label{eq:3pt:lat}
  \sum_{\vec{y}}\, \sum_{\vec{z}}
  \e^{-\I \vec{p}\,'\cdot\vec{y}}\, \e^{\I \vec{q}\cdot\vec{z}} 
  \bret{\Tr\; \Gamma
    G(y,z) \gamma_{\mu} G(z,x) \Gamma^{\dagger} G(x,y)}_g
\end{equation}
with $x_4=0$, $y_4=T/2$, $z_4=t$. Here $G(y,z)$ is the fermion propagator, the
average is taken over the gauge fields, and the trace is over the suppressed
Dirac and colour indices. The matrix $\Gamma$ represents the Dirac structure of
the pion interpolating field $\eta_\pi$, while the Fourier transformations
ensure that we have fixed momenta at the operator insertion and the sink.

We use two different pion interpolating fields to create the pions on the
lattice, namely a pseudo-scalar and the fourth component of the axial-vector
current, which both have the correct quantum numbers. For a given momentum
$\vec{p}$ they read
\begin{equation}
  \label{eq:pions}
  \eta_{\pi}(t,\vec{p}\,) = \sum_{\vec{x}} \e^{-\I \vec{p}\cdot\vec{x}}\;\,
  \overline{d}(x)\, \Gamma\, u(x) \,, \qquad
  \Gamma=\gamma_5\text{ or }\gamma_4\gamma_5
\end{equation}
with $x_4=t$. We apply Jacobi smearing \cite{Allton:1993wc} at the source as
well as the sink to increase the overlap of the lattice interpolating fields
with the physical pion states.

The three-point function~\eqref{eq:3pt:lat} is then evaluated by applying the
sequential source technique as indicated in Fig.~\ref{fig:3pt}. This makes it
efficient to use a large number of momentum transfers, as required for
calculating form factors. A large set of momenta is necessary to assess the
$Q^2$ dependence, and having several combinations of $\vec{p}\,'$ and $\vec{q}$ 
belonging to the same $Q^2$ makes the fits more reliable. We use three final
momenta $\vec{p}\,'$ and 17 momentum transfers $\vec{q}$, giving a total of 51
combinations for an over-constrained fit for $F_\pi$ at 17 different values of
$Q^2$. In units of $2\pi/L$ the momenta are given by
\begin{eqnarray}
\vec{p}\,'\!\!\! &=&  (0,0,0), (0,1,0), (1,0,0), \nonumber \\
\vec{q}    &=&  (0,0,0), (-1,0,0), (-1,-1,0), (-1,-1,-1), \\
           & &  (-2,0,0), (-2,-1,-1), (-2,-2,-1), \cdots
\nonumber 
\end{eqnarray}
where $\cdots$ stands for all possible permutations w.r.t.\ the components.
The errors we quote for our results are statistical errors obtained by the
jackknife method.

\section{Experimental data for the pion form factor\label{sec:experiment}}

Let us now take a brief look at the experimental measurements of $F_\pi(Q^2)$ to
which we compare our lattice results. Very accurate data up to $Q^2=0.253
\gev^2$ have been obtained in \cite{Amendolia:1986wj} from elastic scattering of
a pion beam on the shell electrons of the target material. At higher $Q^2$ the
pion form factor has been extracted from $ep\to en\pi^+$, which is considerably
more involved (see \cite{Blok:2002ew} for a recent discussion). We only use here
data from \cite{Ackermann:1977rp,Brauel:1979zk,Tadevosyan:2006yd+Horn:2006tm},
where the cross sections for longitudinal and transverse photons have been
experimentally separated by the Rosenbluth method.\footnote{For the data from
  \cite{Brauel:1979zk} we use the results of the re-analysis in
  \cite{Volmer:2000ek}} Together these data span a range from $Q^2=0.35 \gev^2$
to $2.45 \gev^2$.

We find the experimental data on $F_\pi$ well described by a monopole form
\begin{equation}
  \label{eq:mono:exp}
F_\pi(Q^2) = \frac{1}{1 + Q^2 /M^2} ,
\end{equation}
with a fit of the combined data from
\cite{Amendolia:1986wj,Ackermann:1977rp,Brauel:1979zk,Tadevosyan:2006yd+Horn:2006tm}
giving $M= 0.714(4)\gev$ at $\chi^2/\text{d.o.f.} = 1.27$. This is remarkably
close to the result $M= 0.719(5)\gev$ at $\chi^2/\text{d.o.f.} = 1.13$ obtained
when fitting only the data of \cite{Amendolia:1986wj} with its much smaller
range in $Q^2$, which illustrates the stability of a monopole form up to
$2.45\gev^2$.

The low-$Q^2$ behaviour of $F_\pi$ is characterised by the squared
charge radius
\begin{equation}
  \label{eq:charge:radius}
\langle r^2 \rangle = - 6\, \left. \frac{\di F_\pi(Q^2)}{\di Q^2} 
  \right|_{Q^2=0} . 
\end{equation}
For a monopole form (\ref{eq:mono:exp}) one has 
\begin{equation}
  \label{eq:mono:radius}
\langle r^2 \rangle = 6 /M^2 .
\end{equation}
In Table~\ref{tab:charge-radius} we list the values obtained from a number of
fits to $F_\pi$. The PDG average \cite{Eidelman:2004wy} uses results from form
factor data at both spacelike and timelike virtualities. The three fits to the
Amendolia data \cite{Amendolia:1986wj} illustrate that different fitting
procedures can give results with a variation much bigger than the quoted
statistical and systematic errors. Fit 1 (whose result is the one retained in
the PDG average) is based on a representation of $F_\pi$ as a dispersion
integral. Fit 2 was also given in \cite{Amendolia:1986wj} and assumed a monopole
form (\ref{eq:mono:exp}) with a normalisation factor allowed to deviate from 1
by $\pm 0.9\%$, which corresponds to the overall normalisation uncertainty of
the measurement. Fit 3 assumes a monopole form with normalisation fixed to 1, as
does the fit to the combined data of
\cite{Amendolia:1986wj,Ackermann:1977rp,Brauel:1979zk,Tadevosyan:2006yd+Horn:2006tm}.
\begin{table}
  \caption{\label{tab:charge-radius} Values of the squared pion charge
    radius obtained from different data sets for $F_\pi(Q^2)$ and with
    different fitting assumptions. Details of the fits are given in the
    text.} 
  \begin{center}
    \renewcommand{\arraystretch}{1.1}
    \begin{tabular}{ll} \hline\hline
      data & $\langle r^2 \rangle$~[fm$^2$]\rule[-1.2ex]{0pt}{2.3ex}\rule{0pt}{2.5ex}\\ \hline
      global average, PDG 2004 \protect\cite{Eidelman:2004wy} & 
      $0.452(11)$\rule{0pt}{2.4ex} \\  
      Amendolia \protect\cite{Amendolia:1986wj}, fit 1 & $0.439(8)$ \\
      \phantom{Amendolia \protect\cite{Amendolia:1986wj},} fit 2 &
      $0.431(10)$ \\ 
      \phantom{Amendolia \protect\cite{Amendolia:1986wj},} fit 3 &
      $0.451(6)$ \\
      combined data
      \protect\cite{Amendolia:1986wj,Ackermann:1977rp,Brauel:1979zk,Tadevosyan:2006yd+Horn:2006tm} ~~ &
      $0.458(5)$ \\ \hline\hline 
    \end{tabular}
  \end{center}
\end{table}

\section{Results\label{sec:res}}
\subsection{Fits to lattice data and extrapolation in $m_\pi$\label{sec:err}}

We start the discussion of our results by explaining our fitting procedure,
including combined fits to all data sets. In the next subsection we will argue
that lattice artifacts are small. To obtain the physical form factor
we have to renormalise our lattice result, $F^{\text{ren}}_{\pi} =
Z_V F^{\text{bare}}_{\pi}$. As mentioned in Section~\ref{sec:lat}, we can
do this by using the electric charge of the pion as input, i.e.\  
\begin{equation}
  \label{eq:norm}
  F^{\text{lat,ren}}_{\pi}(Q^2) =
  \frac{F^{\text{lat,bare}}_{\pi}(Q^2)}{F^{\text{lat,bare}}_{\pi}(0)}\,,
\end{equation}
so that $F^{\text{lat,ren}}_{\pi}(0)=F^{\text{phys}}_{\pi}(0)=1$.  We then use a
monopole ansatz to fit the actual data for the form factor\footnote{We will from
  now on use the renormalised values and drop the superscripts unless required.
  The super- and subscripts `lat' and `phys' respectively refer to observables
  at lattice pion masses and at the physical point.}
\begin{equation}
  \label{eq:mono}
  F^{\text{lat}}_{\pi}(Q^{2})=\frac{1}{1+Q^{2}/M_{\text{lat}}^2}\,,
\end{equation}
where we have $M_{\text{lat}}$ as a fit parameter for each of our lattices at
its lattice pion mass $m_{\pi,\text{lat}}$. The quality of this fitting ansatz
will be discussed below.

Using this fitting function, we compare the results obtained with the two pion
interpolating fields~\eqref{eq:pions} and observe several differences. In
general, the matrix elements for pions using $\Gamma=\gamma_4\gamma_5$ display a
slightly cleaner signal with more data points in $Q^2$, i.e.\ less contamination
due to negative two-point functions. Fitting the monopole form \eqref{eq:mono}
to the form factor for both pion interpolators we find that the
$\chi^2/\text{d.o.f.}$ differs on average by about a factor of 2, ranging from
0.18~--~1.72 (0.23~--~3.49) for the interpolator with $\gamma_{4}\gamma_{5}$
($\gamma_{5}$). The fitted monopole masses for the $\Gamma=\gamma_5$ pions lie
consistently above the ones for $\Gamma=\gamma_4\gamma_5$ but agree within
errors for most lattices. In an exploratory extraction of the pion energies
from the two-point functions with non-vanishing momentum on a sub-set of our
lattices, we also found that the pseudo-scalars with $\Gamma=\gamma_5$ had a
worse signal at higher momenta. A similar observation was made in
\cite{Bonnet:2004fr} and may explain the difference in quality of the form
factors extracted from the two pion currents. Because of the better signal, we
will mainly discuss results for the pions created with $\Gamma =
\gamma_{4}\gamma_{5}$ in the remainder of this work.

\begin{figure}[t]
  \centering
  \includegraphics[angle=270]{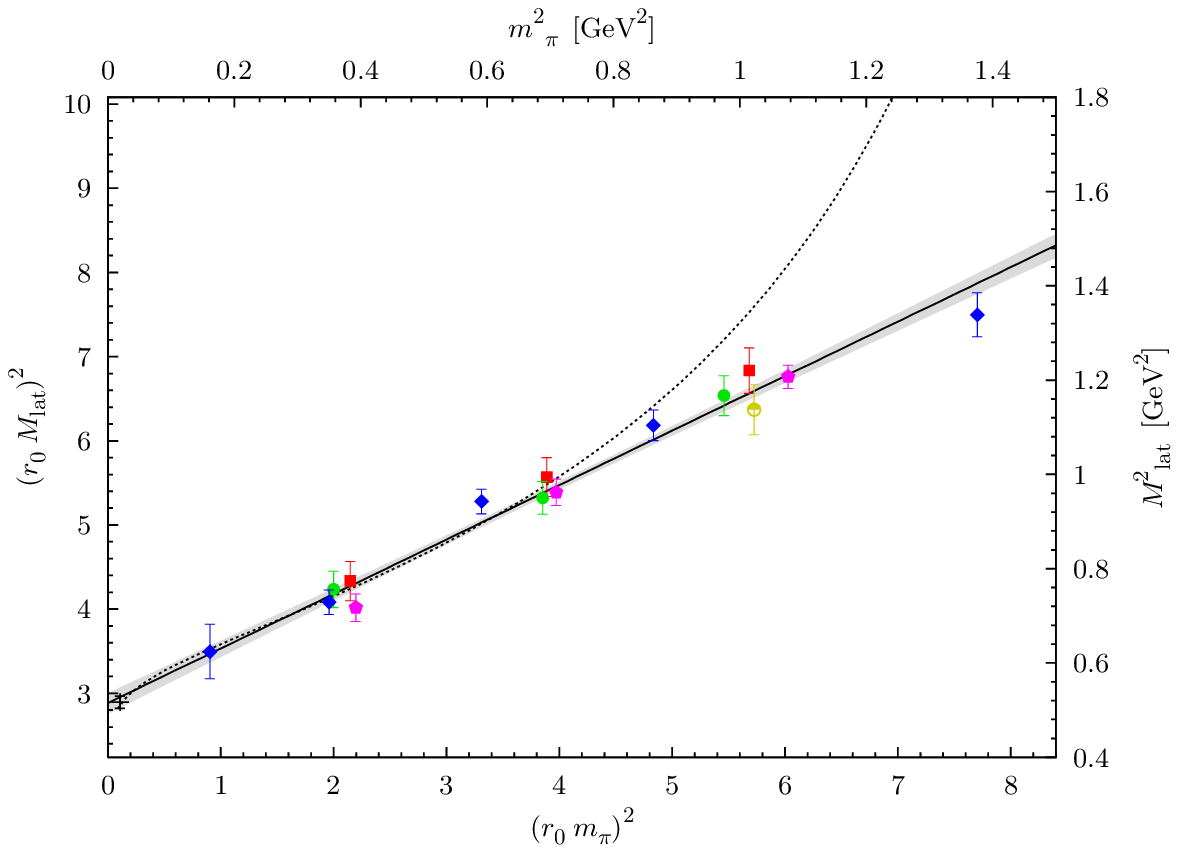}
  \caption{\label{fig:extra} Extrapolations of the squared monopole mass
    against the squared pion mass. The solid line with error band is linear
    extrapolation as obtained from fit 2 while the dotted line shows the central
    curve for fit 4 (whose fit range is limited to $m^2_{\pi} < 0.8\gev^2$). The
    cross marks the monopole mass corresponding to the PDG value
    \protect\cite{Eidelman:2004wy} of the pion charge radius, see
    Eq.~\protect\eqref{eq:mono:radius} and
    Table~\protect\ref{tab:charge-radius}. The different symbols refer to our
    $\beta$-values: squares (5.20), circles (5.25), half-full circle (5.26),
    diamonds (5.29), and hexagons (5.40).}
\end{figure}

To obtain the pion form factor at the physical pion mass we extrapolate the
values for $M_{\text{lat}}$, given in Table~\ref{tab:extra}, to the physical
point. We tried different extrapolations in the square of the pion mass, see
Table~\ref{tab:fit}, including also a fit inspired by chiral perturbation theory
and used in \cite{Hashimoto:2005am}. For the latter we chose the fit range of
$m^2_{\pi, \text{lat}} < 0.8\gev^2$. Varying this fit range within reasonable
bounds did not have a significant effect on the extrapolated value of
$M_{\text{phys}}$. We find the best $\chi^2$ value for fit 2, where
$M_{\text{lat}}^2$ depends linearly on $m_\pi^2$. The extrapolations in the
remainder of this paper are based on this ansatz. We will however include an
estimated systematic error of $\Delta M_{\text{ext}}=35\mev$ from the
difference of fits 1 and 2 in our final result (this is bigger than the
difference between fits 1 and 4, whereas fit 3 gives a significantly worse
description of the data). Figure~\ref{fig:extra} shows the extrapolation to the
physical pion mass based on fits 2 and 4. We remark that our lattice with the
lowest pion mass, $m_\pi=400 \mev$, is completely consistent and increases our
confidence in the fit and fit ansatz.  However, due to the larger statistical
errors it has little weight in this result: when leaving it out of the fit
$M_{\text{phys}}$ changes only by $1\mev$. The corresponding run and several
others at small pion masses are still in progress. It is obvious that one needs
higher statistics for this point to be significant.
\begin{table}[thbp]
  \caption{\label{tab:extra} Monopole masses $M_{\text{lat}}$ obtained
  from fits to \protect\eqref{eq:mono} for each of our lattices.  The
  last column gives an estimate for the shift $\Delta M_{\text{lat}} =
  M(m_\pi^2,\infty) - M(m_\pi^2,L)$ of the monopole mass due to finite
  volume effects. It is obtained from the empirical fit \eqref{eq:fit:vol}
  discussed in Section~\protect\ref{sec:artifacts}.}
  \begin{center}
  \begin{tabular}{cccccr@{}l}
    \hline\hline%
    \# & $m_{\pi}$~[\Gev] & $L$~[fm]& $m_{\pi} L$ & $M_{\text{lat}}~[\Gev]$ &
    \multicolumn{2}{c}{$\Delta
      M_{\text{lat}}~[\Mev]$\rule{0pt}{2.3ex}\rule[-1.2ex]{0pt}{2.5ex}}\\%
    \hline
   1 & 1.007(2) & 1.8 & 9.4 & 1.104(22) & 0&.3 \rule{0pt}{2.4ex}\\
   2 & 0.833(3) & 1.6 & 6.6 & 0.997(21) & 4&.3\\%
   3 & 0.619(3) & 1.5 & 4.7 & 0.880(24) & \hspace{1em}35&.2\\%
\hline                                  
   4 & 0.987(2) & 1.6 & 7.9 & 1.089(20) & 1&.1\\%
   5 & 0.829(3) & 1.5 & 6.1 & 0.975(17) & 7&.2\\%
   6 & 0.597(1) & 2.0 & 6.1 & 0.870(22) & 8&.0\\%
\hline                                  
   7 & 1.011(3) & 1.6 & 8.1 & 1.066(25) & 0&.9\\%
\hline                                  
   8 & 1.173(2) & 1.6 & 9.2 & 1.157(20) & 0&.3\\%
   9 & 0.929(2) & 1.4 & 6.7 & 1.051(15) & 3&.7\\%
   10& 0.769(2) & 2.0 & 7.8 & 0.971(14) & 1&.3\\%
   11& 0.591(2) & 1.9 & 5.7 & 0.854(15) & 12&.6\\%
   12& 0.400(3) & 1.9 & 3.8 & 0.783(36) & &--\\%
\hline                                  
   13& 1.037(1) & 1.8 & 9.7 & 1.099(13) & 0&.2\\%
   14& 0.842(2) & 1.8 & 7.5 & 0.981(14) & 1&.8\\%
   15& 0.626(2) & 1.7 & 5.3 & 0.847(17) & 18&.2\\%
    \hline\hline
  \end{tabular}
  \end{center}
\end{table}
\begin{table}[htp]
  \caption{Different forms used to extrapolate the monopole mass to
  the physical value of $m_\pi$. In fit 4 we have $L = 1/(4\pi f)^2
  \log(m^2_{\pi, \text{lat}}/\mu^2)$, where $\mu=1\gev$ and $f_{\pi}\approx
  92\mev$ is the pion decay constant.}
  \label{tab:fit}
  \begin{center}
    \renewcommand{\arraystretch}{1.2}
  \begin{tabular}{llcr@{}lr@{.}l}
    \hline\hline
    \# & extrapolation ansatz & $\chi^2$/d.o.f.\ & \multicolumn{2}{c}{$c_1$} &
    \multicolumn{2}{c}{$M_{\text{phys}}$
      [\Gev]\rule{0pt}{2.5ex}\rule[-1.3ex]{0pt}{2.5ex}}\\\hline
    1 & $M_{\text{lat}}^{\phantom{2}} = c_0 + c_1 m^2_{\pi, 
         \text{lat}}$\rule{0pt}{2.3ex} & 1.31 & 0.322(15)&$\gev^{-1}$ &
       0&761(13)\\
    2 & $M_{\text{lat}}^2 = c_0 + c_1 m^2_{\pi, 
         \text{lat}}$ & 0.93 & 0.647(30)& & 0&726(16)\\
    3 & $1/M_{\text{lat}}^2 = c_0 + c_1 m^2_{\pi, 
         \text{lat}}\rule[-1.1ex]{0pt}{2.5ex}$ & 3.25 & $-0.575(31)$&$\gev^{-4}$
       & \hspace{.7em}0&833(9)\\
    4 & $6/M_{\text{lat}}^2 = c_0 + c_1 m^2_{\pi,\text{lat}} - L$ & 1.11 &
    $-4.33(62)$&$\gev^{-4}$ & 0&715(4)\\
    \hline\hline
  \end{tabular}
  \end{center}
\end{table}

We include the $m_{\pi}$ dependence of the monopole mass of fit 2 in a combined
fit to all our lattice data available. This fit has the same monopole form as in
\eqref{eq:mono} with one additional parameter to incorporate the $m_{\pi}$
behaviour,
\begin{equation}
  \label{eq:fit:global}
  \begin{split}
    F_{\pi}(Q^2,m_\pi^2) &= \frac{1}{1+Q^{2}/M^2(m_\pi^2)}\,, \\[0.4em]
    M^2(m_\pi^2) &= c_0 + c_1 m^2_{\pi}\,.
  \end{split}
\end{equation}
The two fit parameters, $c_0$ and $c_1$, describe the relation between the
monopole mass and the pion mass, and we immediately obtain the form factor
$F^{\text{phys}}_\pi(Q^2) = F^{\phantom{h}}_\pi(Q^2,m_{\pi,\text{phys}}^2)$ in
the physical limit.  The fitted parameters are $c_0=0.517(23)\gev^2$ and
$c_1=0.647(30)$ with $\chi^2/\text{d.o.f.}=0.64$. This gives
$M_{\text{phys}}=M(m^2_{\pi,\text{phys}})=0.727(16)\gev$, in good agreement with
the experimental result.

\begin{figure}[t]
  \centering
  \includegraphics[angle=270]{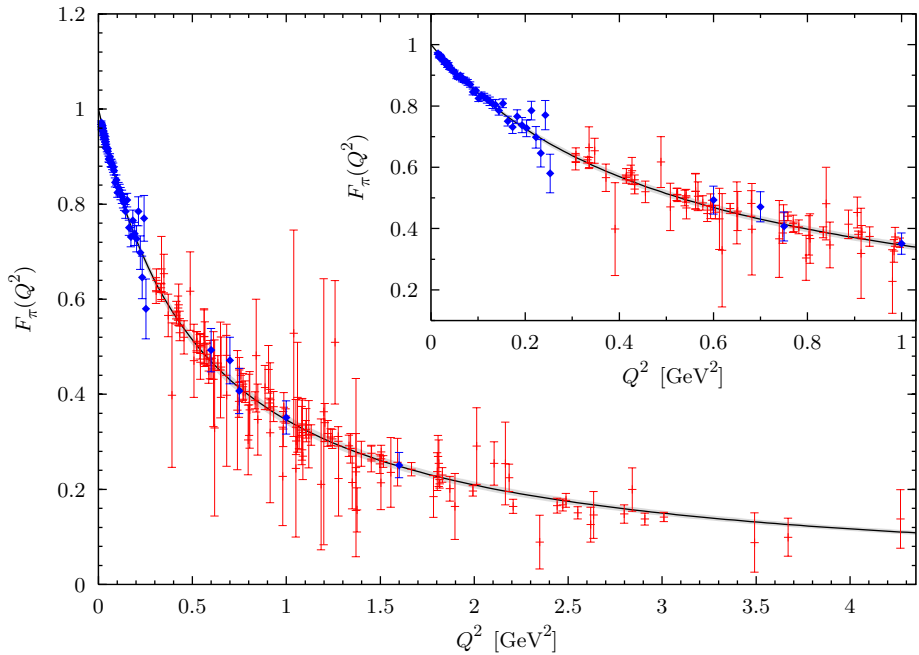}
  \caption{\label{fig:global} Combined fit to~\eqref{eq:fit:global} of
    our data for all lattices. We plot experimental data (diamonds)
    \protect\cite{Amendolia:1986wj,Brauel:1979zk,Volmer:2000ek} and lattice
    results extrapolated to the physical pion mass as explained in the
    text. To avoid having a cluttered plot we do not show lattice
    results with errors bigger than 80\%, which are nevertheless
    included in the fit. The insert shows the good agreement to the
    experimental data for a momentum transfer of up to 1\gev$^2$. Also
    included is an error band for the fit.}
\end{figure}

Figure \ref{fig:global} shows experimental data along with the combined fit with
its extrapolated curve. For this plot, our data at the lattice pion masses is
shifted to the physical pion mass and plotted on-top of the extrapolation. We do
this by subtracting from the individual lattice points,
$F^{\text{lat}}_{\pi}(Q^2)$, a value $\bigr( F_{\pi}(Q^2,m_{\pi,\text{lat}}^2) -
F_{\pi}(Q^2,m_{\pi,\text{phys}}^2) \bigr)$ calculated with the fit parameters of
Eq.~\eqref{eq:fit:global} at the respective pion masses. The errors are left
unchanged. We find good agreement between our simulation and the experimental
results. This is emphasised by the insert in Fig.~\ref{fig:global}, which shows
the region $Q^2<1\gev^2$, where most of the experimental points lie. The same
fit for the pions with $\Gamma = \gamma_5$ gives
$M_{\text{phys}}=0.773(17)\gev$, with a bigger $\chi^2/\text{d.o.f.}$ of 1.01.

\begin{figure}[tbp]
  \centering
  \includegraphics{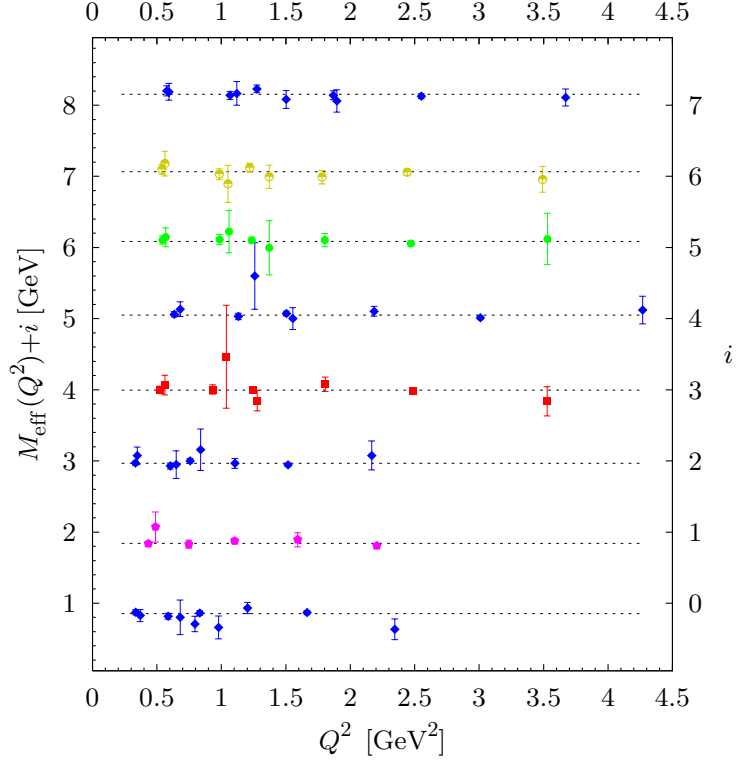}%
  \caption{\label{fig:test1} Effective monopole masses
    $M_{\text{eff}}(Q^2)$ defined in \protect\eqref{eq:eff:mass}, together with
    the corresponding monopole masses from Table~\ref{tab:extra} (dotted lines)
    for a sample of our lattices from small to large pion masses (lattices
    number 8, 7, 4, 9, 2, 10, 15, and 11 from top to bottom). For better
    visibility we omitted 2 points with very large errors in the plot, but
    included them in the fit.}
\end{figure}

\begin{figure}[tbp]
  \centering
  \includegraphics[angle=-90]{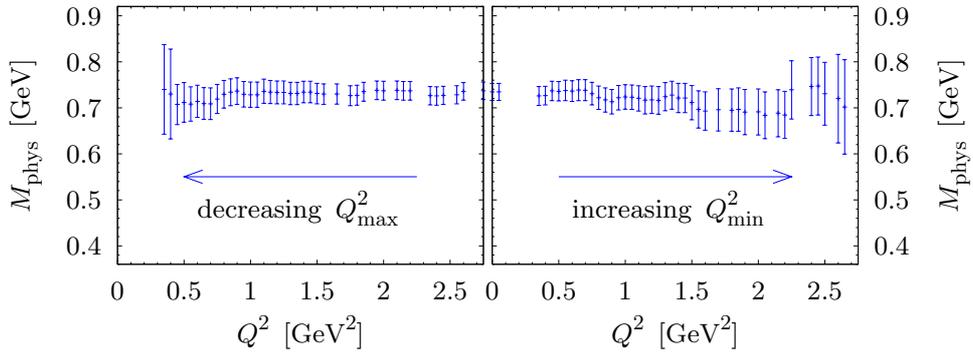}
  \caption{\label{fig:test2} Combined fits to
    \protect\eqref{eq:fit:global} with reduced fitting ranges in
    $Q^2$. For the left plot $Q^2_{\text{max}}$ is decreased, while
    $Q^2_{\text{min}}$ is increased for the right plot. We use bins of
    $50\mev^2$ and show only points where the number of data points in
    the fit of $F_\pi$ changed.}
\end{figure}

We now investigate the validity of the monopole ansatz for our data. Instead of
constraining the fitting function to a monopole form, one can also take a
general power law, i.e.\ use a function
\begin{equation}
  \label{eq:fit:pole}
  \begin{split}
    F_{\pi}(Q^2,m_\pi^2) &= 
      \left(1+\frac{Q^{2}}{p\slim M^2(m_\pi^2)}\right)^{-p}\,, \\[0.4em]
    M^2(m_\pi^2) &= c_0 + c_1 m^2_{\pi}\,,
  \end{split}
\end{equation}
with an additional parameter, $p$. Note that the relation \eqref{eq:mono:radius}
is still valid, independent of $p$. A combined fit to all our data sets results
in $p=1.173(58)$, now with a mass $M_{\text{phys}}=0.757(18)\gev$ and a
$\chi^2/\text{d.o.f.}=0.58$, indicating that the monopole form is a good
description. Taking the difference between this number and the result of the fit
to \eqref{eq:fit:global}, we assign a systematic error of $\Delta
M_{\text{fit}}=30\mev$ to $M_{\text{phys}}$ due to the ansatz for the fitting
function. Another alternative is to calculate an effective monopole mass for
every momentum $Q^2$ separately by solving Eq.~\eqref{eq:mono} for
$M_{\text{lat}}$:
\begin{equation}
  \label{eq:eff:mass}
  M_{\text{eff}}(Q^2) = Q \left[ \frac{1}{F^{\text{lat}}_{\pi}(Q^2)} - 1\,
  \right]^{-1/2} .
\end{equation}
We show such effective masses for some of our lattices in Fig.~\ref{fig:test1},
where one can see that the effective monopole masses stay constant within errors
over a large range of $Q^2$ and agree with the monopole masses given in
Table~\ref{tab:extra}. This again indicates that the monopole is a good
description for our data. The validity of the fit over the whole $Q^2$ range is
further tested by combined fits to Eq.~\eqref{eq:fit:global} in a limited
fitting range $Q^2\le Q^2_{\text{max}}$ or $Q^2_{\text{min}} \le Q^2$. This is
shown in Fig.~\ref{fig:test2}, where we successively limit the fit to smaller
(larger) momenta. Note that the increasing errors to the left or the right are
due to the decrease in the number of fitted data points. Within these errors,
the change in the monopole mass is consistent with statistical fluctuations.
From Figs.~\ref{fig:test1} and \ref{fig:test2} we can conclude that the monopole
ansatz works well in the entire region for which we have lattice data, from
$Q^2=0$ to about $4~\Gev^2$.


The results discussed so far have used the lattice data normalised as in
\eqref{eq:norm}. Using
\begin{equation}
  \label{eq:zv}
  Z_V F^{\text{lat,bare}}_{\pi}(0) = F_{\pi}^{\text{lat,ren}}(0) = 1,
\end{equation}
we can determine $Z_V$ from our (unrenormalised) data at zero momentum transfer.
We find reasonable agreement with the values of $Z_V$ given in
\cite{Bakeyev:2003ff}, albeit with errors that are larger by at least an order
of magnitude. The bigger errors are likely due to our choice of $\tsi$,
which results in noisier two-point functions.

\subsection{Finite volume and discretisation effects}
\label{sec:artifacts}

\begin{figure}[tb]
  \centering
  \includegraphics[angle=270]{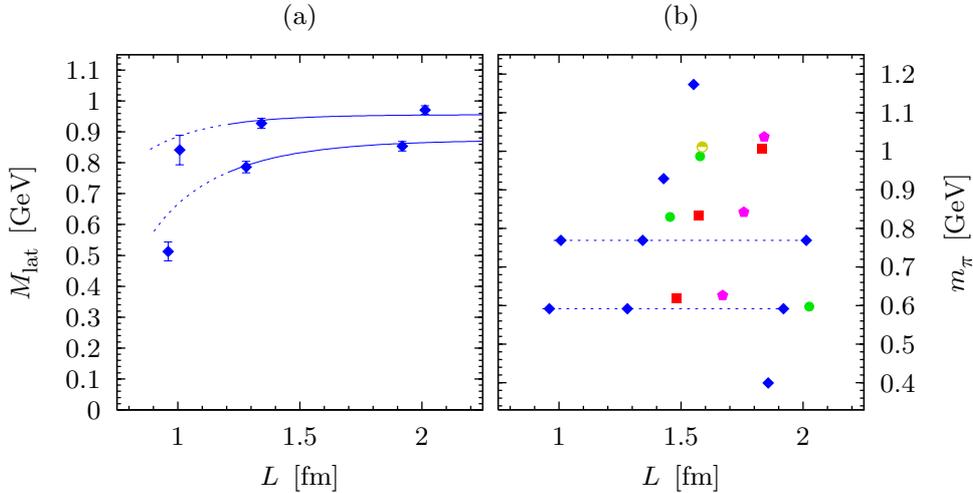}
  \caption{\label{fig:vol} (a) Monopole mass vs.\ lattice size in our
    finite volume data sets with $\beta=5.29$ and $\kappa=0.1355$
    (upper points) or $\kappa=0.1359$ (lower points).  The curves
    correspond to a fit to \protect\eqref{eq:fit:vol} as discussed in
    the text.  (b) Overview of pion masses and lattice sizes for our
    complete data set. The dotted lines mark our finite size runs.}
\end{figure}

Let us now turn to the discussion of lattice artifacts. Apart from the
extrapolation to the physical pion mass there are two more limits to be taken:
the infinite volume limit and the continuum limit. The large number of lattices
available allows us to investigate both.  In order to study the volume
dependence of our results, we make use of two sets of configurations that have
the same parameters $\beta$, $\kappa$ for the lattice action but different
volumes (see Table~\ref{tab:lats2}).
\begin{table}[bt]
  \caption{\label{tab:lats2} Overview of our finite size runs. Note that we use
    the pion mass and lattice spacing of the largest lattice also for the
    smaller ones. They are given in Table~\ref{tab:lats} and not repeated here.}
  \begin{center}
  \begin{tabular}{cclccccr@{}l}
    \hline\hline%
    $\beta$ & $\kappa$ & \hspace{.7ex}\# & $N^3\times T$ & $L$~[fm]& $m_{\pi}L$ &
    $M_{\text{lat}} [\Gev]$ & \multicolumn{2}{c}{$\Delta M_{\text{lat}} [\Mev]$\rule{0pt}{2.3ex}\rule[-1.2ex]{0pt}{2.5ex}}\\%
    \hline
   5.29 & 0.13550 & 10& $24^3\times 48$& 2.0 & 7.8 & 0.971(14) & 1&.4
   \rule{0pt}{2.4ex}\\%
        &&10$\slim$a& $16^3\times 32$  & 1.3 & 5.2 & 0.928(16) & 19&.7\\%
        &&10$\slim$b& $12^3\times 32$  & 1.0 & 3.9 & 0.841(48) & 75&.0\\\hline%
   5.29 & 0.13590 &11& $24^3\times 48$ & 1.9 & 5.7 & 0.854(15) & 12&.6
        \rule{0pt}{2.4ex}\\%
        &&11$\slim$a& $16^3\times 32$  & 1.3 & 3.8 & 0.786(18) & 90&.3\\%
        &&11$\slim$b& $12^3\times 32$  & 1.0 & 2.9 & 0.513(31) &
        \hspace{1em}263&.1\\%
    \hline\hline
  \end{tabular}
  \end{center}
\end{table}
In Fig.~\ref{fig:vol}a we show the monopole masses fitted according to
Eq.~\eqref{eq:mono} as a function of the lattice size $L$. We use the pion mass
$m_{\pi}$ and lattice spacing $a$ determined for the lattice with the largest
volume also for the smaller ones. Figure~\ref{fig:vol}b gives an overview of our
lattices in the $m_{\pi}$--$L$ plane.

To obtain some understanding of the volume dependence one may have recourse to
chiral perturbation theory. The volume dependence of the pion charge radius has
been investigated to one-loop order in various approaches of chiral perturbation
theory \cite{Bunton:2006va,Chen:2006gg,Borasoy:2004zf}. In the continuum limit,
the result of the lattice regularised calculation in \cite{Borasoy:2004zf}
amounts to a finite size correction of
\begin{equation}
  \label{eq:chiralfse}
  \bret{r^2}_L - \bret{r^2}_{\infty} = \frac{3}{8 \pi^2 f^2_{\pi}} \sum_{\vec{n}
    \neq \vec{0}} K_0 (L m_\pi |\vec{n}|)\,,
\end{equation}
where the sum runs over all three-vectors $\vec{n} \neq \vec{0}$ with integer
components and $f_{\pi}\approx 92\mev$ is the pion decay constant. Note that the
finite size correction of the charge radius is not proportional to $m^2_{\pi}$,
unlike for other quantities such as the pion decay constant or the nucleon axial
coupling. The leading contribution in Eq.~\eqref{eq:chiralfse} for large values
of $m_{\pi} L$ is proportional to $K_0 (m_{\pi} L) \sim \sqrt{\pi/(2 m_{\pi}
  L)}\, \e^{-m_{\pi} L}$. Unfortunately we cannot expect chiral perturbation
theory to be applicable at the pion masses and lattice volumes used in our
simulations. This includes the result \eqref{eq:chiralfse}, which we take
however as a guide for the functional form of the volume dependence. We thus
change the monopole mass in \eqref{eq:fit:global} to\footnote{Taking the Bessel
  function $K_0 (m_{\pi} L)$ instead of $\e^{-m_{\pi}L}$ does not change our
  results significantly.}
\begin{equation}
  \label{eq:fit:vol}
  M^2(m_\pi^2,L) = c_0 + c_1 m^2_{\pi} + c_2\slim \e^{-m_{\pi}L}\,.
\end{equation}
We then perform a combined fit to the data of all lattices in
Table~\ref{tab:lats} except for number $12$ (see below), including in addition
the $16^3\times32$ lattices of the finite volume runs (numbers 10$\slim$a and
11$\slim$a). The result is represented by the solid lines in
Fig.~\ref{fig:vol}a. The fitted parameters are $c_0=0.553(29)\gev^2$,
$c_1=0.612(35)$ and $c_2=-6.97(1.71)\gev^2$ at $\chi^2/\text{d.o.f.}=0.62$ which
gives $M_{\text{phys}}=0.751(19)\gev$ for the infinite volume limit of the
monopole mass at the physical point. Compared with the value $0.727(16)\gev$
obtained in the fit \eqref{eq:fit:global} without volume dependence, this
represents a small overall finite-size effect. The fitted parameters do not
change significantly if we only fit the $16^3\times32$ and $24^3\times48$ data
sets of the finite volume runs, i.e.\ the data corresponding to the four
rightmost points in Fig.~\ref{fig:vol}a (lattices number 10, 10$\slim$a, 11, and
11$\slim$a). We have not included the $12^3\times32$ lattices in the fit
\eqref{eq:fit:vol} since we cannot expect our simple ansatz to hold down to
lattice sizes of 1~fm.  Qualitatively, our fit is not too bad even in this
region, as shown by the dotted lines in Fig.~\ref{fig:vol}a.

With the fitted parameters we can estimate the finite volume shift for each of
our lattices as given in Table~\ref{tab:extra}. Except for a few lattices we
find very small effects. We do not expect that with the simple form
\eqref{eq:fit:vol} fitted to our finite volume data at $m_{\pi}=591\mev$ and
$m_{\pi}=769\mev$ (the dotted lines in Fig.~\ref{fig:vol}b) we can estimate
volume effects for pion masses as low as $400\mev$. We therefore have excluded
lattice number $12$ from our finite volume investigation.

\begin{figure}[!bt]
  \centering
  \includegraphics[angle=270]{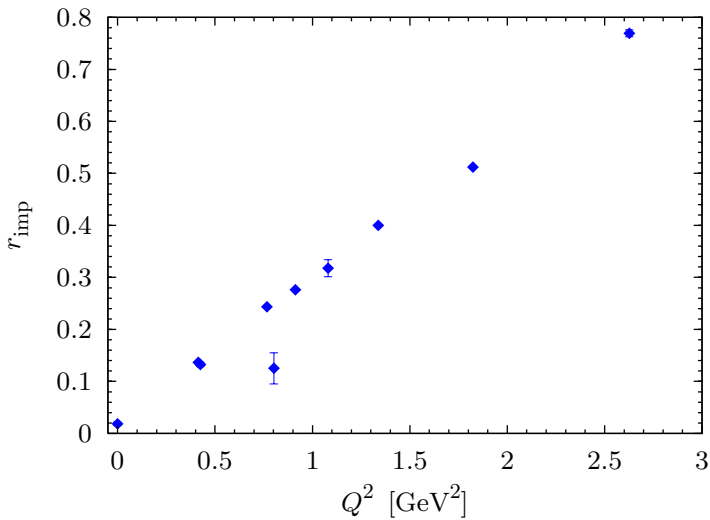}
  \caption{\label{fig:impro} The ratio $r_{\text{imp}}(Q^2)$ defined
  in Eq.~\protect\eqref{eq:impro:ratio}, evaluated for our coarsest
  lattice ($\beta=5.20$, $\kappa=0.1342$). To obtain the effect of
  $\op{O}(a)$ improving the current, this ratio needs to be multiplied
  with $c_V$.}
\end{figure}

Before discussing the scaling behaviour, let us briefly discuss the possibility
of $\op{O}(a)$ improving the local vector current. The improved
current has the form
\begin{equation}
  \label{eq:impro:current}
\begin{split}
  V^{\text{imp}}_{\mu}(x) &= \overline{u}(x) \gamma_{\mu} u(x) + c_V
  \slim a \partial_{\nu} 
  T_{\mu\nu}(x) \,, \\[0.2em]
  T_{\mu\nu}(x) &= \I\, \overline{u}(x) \sigma_{\mu\nu} u(x)\,.
\end{split}
\end{equation}
The improvement coefficient $c_V$ is only known from lattice perturbation theory
\cite{Sint:1997jx} because the only non-perturbative calculations to date are
for quenched fermions (see e.g.\ \cite{Guagnelli:1997db}). However, even with
tadpole improvement the perturbative value for our coarsest lattice is $c_V
\approx -0.027$. This is so small that we expect no sizable effect on our
results. To see this, we plot in Fig.~\ref{fig:impro} the ratio
\begin{equation}
  \label{eq:impro:ratio}
  r_{\text{imp}}(Q^2) = \frac{\bra{\pi(p')} {a \partial_{\nu} 
      T_{\mu\nu}} \ket{\pi(p)}}{\bra{\pi(p')} {\overline{u}
      \gamma_{\mu} u} \ket{\pi(p)}}
\end{equation}
of the pion matrix elements for the two operators on the r.h.s.\ of
Eq.~\eqref{eq:impro:current}. The dependence on the index $\mu$ cancels in this
ratio. Note that here we use unrenormalised lattice data and that we still have
to multiply with $c_V$ in order to obtain the effect of the improvement term in
the current. This example plot is for our coarsest lattice ($\beta=5.20$ and
$\kappa=0.1342$), where the improvement term should have the largest impact.  To
gain a feeling for the possible size of the effect, we used a fixed value of
$c_V=-0.3$ to compute the effect on a sub-set of our lattices (lattices number
2, 6, 11, 15). Although this improvement coefficient is more than ten times
larger than the tadpole improved value for our coarsest lattice, the shift of
the monopole mass was moderate with 6 to 10\%. Given the size of our statistical
errors on $F_\pi$ and the fact that a reliable value for $c_V$ is not known for
our lattices, we decided to neglect operator improvement and use the local
vector current.

\begin{figure}[!tb]
  \centering
  \setlength{\unitlength}{1cm}
  \includegraphics[angle=270]{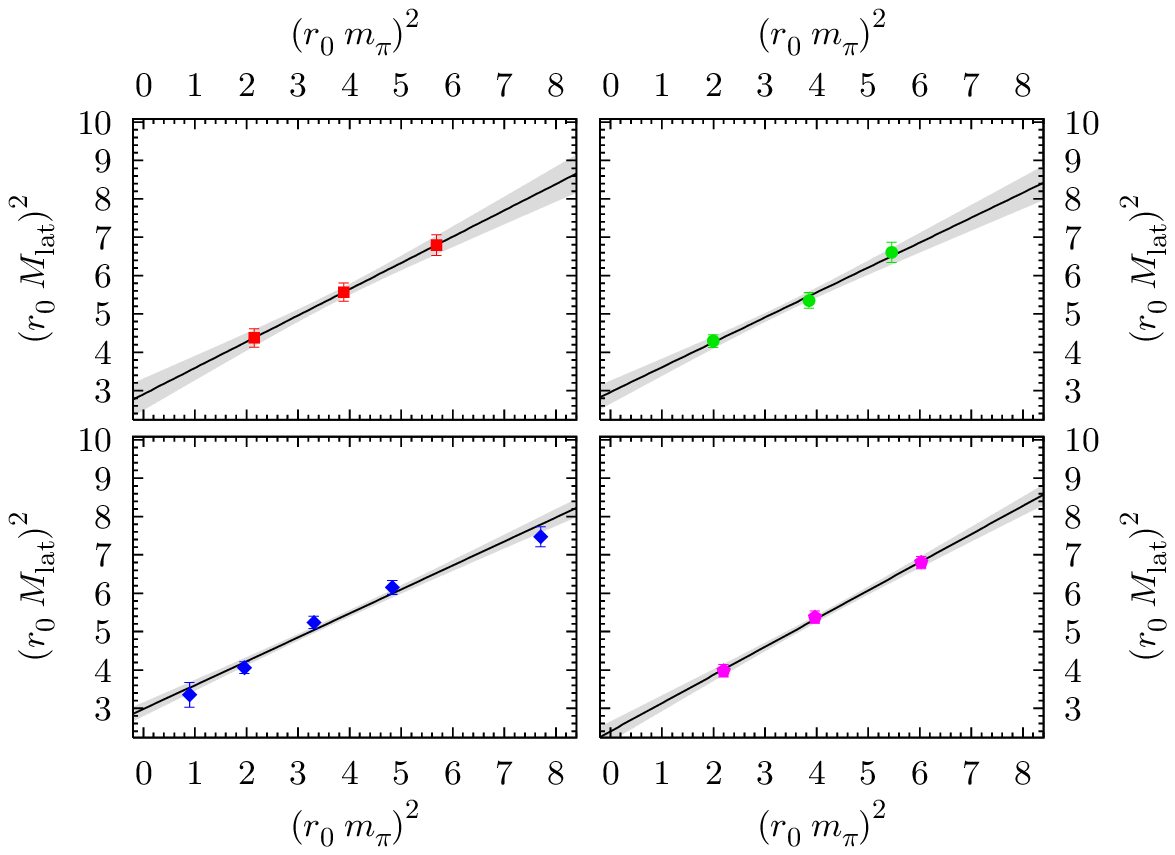}
  \begin{picture}(12,0)
    \put(1.65,6.8){\makebox{$\beta=5.20$}}
    \put(6.5 ,6.8){\makebox{$\beta=5.25$}}
    \put(1.65,3.6){\makebox{$\beta=5.29$}}
    \put(6.5 ,3.6){\makebox{$\beta=5.40$}}
  \end{picture}
  \includegraphics[angle=270]{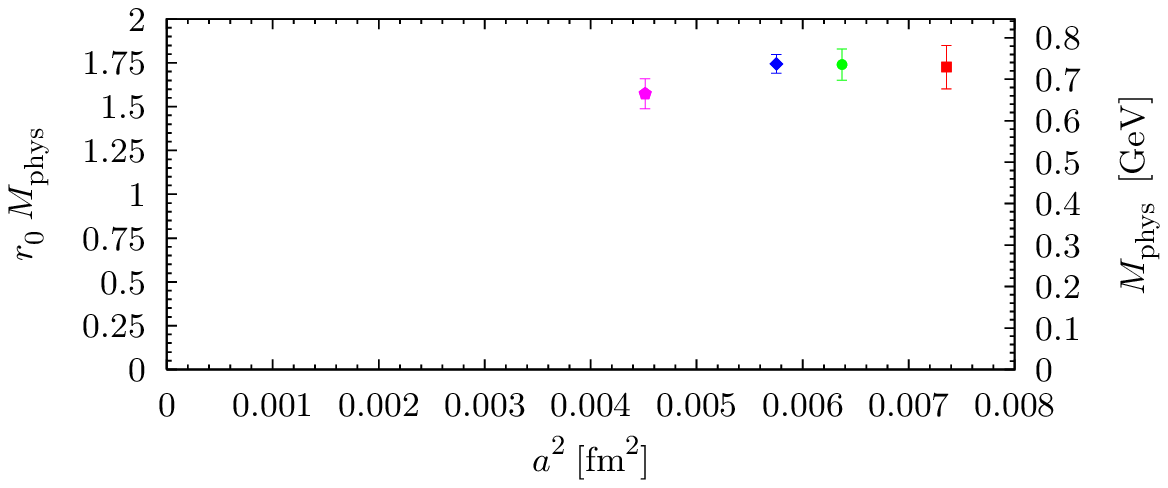}
  \caption{\label{fig:sca1} Scaling test: the upper plots show
    extrapolations as in Fig.~\ref{fig:extra} for each $\beta$ separately. The
    lower plot shows the extrapolated values of the monopole mass at the
    corresponding square of the lattice spacing.}
\end{figure}

We now investigate the scaling behaviour by extrapolating our values for the
monopole mass to the physical pion mass separately for each $\beta$ (see the
upper plots in Fig.~\ref{fig:sca1}). We again assume a linear relation between
the squared monopole and pion masses. The extrapolated values can then be
studied as a function of the lattice spacing $a$, using $r_0/a$ extrapolated to
the chiral but not to the continuum limit \cite{Gockeler:2005rv}.\footnote{We
  have updated values for $r_0/a$ w.r.t.\ \cite{Gockeler:2005rv}: for
  $\beta=5.20, 5.25, 5.29, \text{and }5.40$ they are $r_0/a = 5.444(72),
  5.851(85), 6.158(53), \text{and } 6.951(54)$, respectively.} This is shown
in the lower plot in Fig.~\ref{fig:sca1}. While the three
  rightmost data points in the lower plot of Fig.~\ref{fig:sca1} strongly suggest
  that no discretisation errors are present within statistical errors, it
  requires additional simulation points to see if the leftmost data point in the
  lower plot of Fig.~\ref{fig:sca1} represents a downwards trend or is just
an outlier.
From the discussion above and the overview in Table~\ref{tab:extra} we recall
that some of the points at low pion mass may be affected by finite volume
corrections. We have repeated the fits shown in Fig.~\ref{fig:sca1} with squared
monopole masses shifted upwards by $c_2\slim \e^{-m_{\pi}L}$, where $c_2$ (in
units of $r_0^{-2}$) was taken from the global fit described after
Eq.~\eqref{eq:fit:vol}. Note that the pion mass of $400\mev$ is excluded from
this global fit for the reasons given above. The result shows an increase of
$M_{\text{phys}}$ mainly for $\beta=5.20$ and $5.40$ but is again consistent
with no $a$ dependence. Given the lever arm in $a^2$ and the size of our
statistical and finite size errors, we refrain from including an explicit $a$
dependence of the monopole mass in our global fit \eqref{eq:fit:global}.

\section{Conclusion\label{sec:conclusio}}

We have calculated the electromagnetic form factor of the pion, using lattice
configurations generated by the QCDSF/UKQCD collaboration with two flavours of
dynamical, $O(a)$ improved Wilson fermions. The corresponding pion masses range
from 400 to 1180~$\mev$.  The momentum dependence of the pion form factor was
studied up to $Q^2$ around $4\gev^2$. Within errors, the pion form factor is
described very well by a monopole form \eqref{eq:mono} in this range, for all
our lattice pion masses. A linear chiral extrapolation to the physical pion mass
leads to a monopole mass of $M=0.727(16)\gev$. This corresponds to a squared
charge radius $\langle r^2 \rangle = 0.441(19)~\text{fm}^2$, in good agreement
with experiment. Our extrapolated lattice data for the form factor is compared
with experimental measurements in Fig.~\ref{fig:global}. Other lattice results
are quoted in Table~\ref{tab:others}.

The large parameter space of the gauge configurations we used makes it possible
to explore artifacts arising from the finite lattice spacing and volume.  An
empirical fit allowing for a volume dependence leads to an increase of the
monopole mass by 3\% at infinite volume and the physical point. Within errors,
our results show no clear dependence on the lattice spacing in the range
$a=\text{0.07~--~0.11~fm}$ of our simulations.  Including estimates for
systematical errors, our final result then is $M=0.727\pm 0.016\,(\text{stat})
\pm 0.046\,(\text{syst}) + 0.024\,(\text{vol})\gev$, which translates to a
charge radius of $\langle r^2 \rangle = 0.441\pm 0.019\,(\text{stat}) \pm
0.056\,(\text{syst}) -0.029\,(\text{vol})~\text{fm}^2$. The first error is
purely statistical, followed by a systematic uncertainty due to the ansatz for
the fitting function and the extrapolation to physical pion masses (for which we
added in quadrature the errors $\Delta M_{\text{ext}}$ and $\Delta
M_{\text{fit}}$ obtained in Section~\ref{sec:err}). The last error reflects a
possible shift because of finite volume effects as just discussed. We have set
the scale using the Sommer parameter with $r_0=0.467$~fm. We note that the
analysis leading to our result for $M$ is independent of the scale setting, so
that a different value of $r_0$ would lead to a simple rescaling of the above
values.
\begin{table}[tp]
  \caption{An overview of lattice results for the pion charge radius along
    with the experimental value. We only quote results that are extrapolated to
    the physical point. The quoted lattice errors are purely statistical.}
  \label{tab:others}
  \begin{center}
    \renewcommand{\arraystretch}{1.1}
  \begin{tabular}{lll}
    \hline\hline
    $\langle r^2 \rangle$~[fm$^2$] & type of result &
    Reference\rule{0pt}{2.5ex}\rule[-1.1ex]{0pt}{2.3ex}\\\hline
    0.452(11) & experimental value & PDG 2004
    \cite{Eidelman:2004wy}\rule{0pt}{2.3ex}\\\hline
    0.441(19) & Clover improved Wilson fermions, $N_f=2$ & this
    work\rule{0pt}{2.3ex}\\
    0.396(10) & Clover improved Wilson fermions, $N_f=2$ & JLQCD
    \cite{Hashimoto:2005am}\\
    0.37(2) & Wilson fermions, quenched &\cite{vanderHeide:2003kh}\\
    0.310(46) & hybrid ASQTAD/DWF, $N_f=2+1,3$ & LHPC \cite{Bonnet:2004fr}\\
    \hline\hline
  \end{tabular}
  \end{center}
\end{table}

\section*{Acknowledgements}

The numerical calculations have been performed on the Hitachi SR8000 at LRZ
(Munich), on the Cray T3E at EPCC (Edinburgh) \cite{Allton:2001sk}, and on the
APEmille and apeNEXT at NIC/DESY (Zeuthen). The simulations at the smallest pion
mass have been done on the Blue Gene/L at NIC/J\"ulich and by the Kanazawa group
on the Blue Gene/L at KEK as part of the DIK research programme. This work is
supported in part by the DFG (Forschergruppe Gitter-Hadronen-Ph\"anomenologie),
by the EU Integrated Infrastructure Initiative ``Hadron Physics'' (I3HP) under
contract number RII3-CT-2004-506078, and by the Helmholtz Association under
contract number VH-NG-004. D.B.\ thanks the School of Physics, University of
Edinburgh, where this work was finished, for its hospitality. Ph.H.\ 
acknowledges support by the DFG Emmy-Noether programme.

\end{document}